\documentclass[
reprint,
 amsmath,amssymb,
 aps, 
pra,
floatfix,
showkeys
]{revtex4-2}

\usepackage{graphicx}
\usepackage{dcolumn}
\usepackage{bm}
\usepackage{physics}
\usepackage{placeins}
\usepackage{subscript}
\usepackage{appendix}
\usepackage[utf8]{inputenc}

\begin{document}

\preprint{APS/123-QED}

\title{\textbf{Ab-initio study of high harmonic generation from fullerenes: Multi-orbital effects, correlations, and size dependence}}
%
%

\author{Km Akanksha Dubey}
\email{akankshadubey256@gmail.com}
\affiliation{Technion - Israel Institute of Technology, Schulich Faculty of Chemistry, Haifa, 32000036, Israel}
\author{Ofer Neufeld}%
\email{ofern@technion.ac.il}
\affiliation{Technion - Israel Institute of Technology, Schulich Faculty of Chemistry, Haifa, 32000036, Israel}%


\begin{abstract}
 We study with \textit{ab-initio} theory high harmonic generation (HHG) from a series of gas-phase fullerenes (from C\textsubscript{20} to C\textsubscript{60}, including isomers) under varying laser conditions (linearly polarized at various angles and different ellipticities). We explore the role of multiple orbitals in the dynamics and show that due to the tight energy level spacing in these systems (forming the equivalent of energy bands), multiple orbitals contribute and cause severe spectral interferences. HHG cutoff and plateau ranges are shown relatively similar across species of different sizes. We further explore the role of correlations, which are known to be prevalent in C$_{60}$ and similar species and responsible for plasmonic resonances. We find that the independent-particle approximation, where electrons are not dynamically interacting, fails to quantitatively describe features in the spectrum besides the cutoff energy. This failure is broad across the HHG plateau and perturbative regime in all species. Broadly, correlations are seen to reduce HHG yields and cause sharper emission peaks. Lastly, we study HHG ellipticity and angular dependence across species and show that: (i) HHG angular sensitivity becomes less prevalent for larger fullerenes. This arises from a more isotropic structure of the higher point-group molecules. (ii) HHG yields decay faster with the driving laser ellipticity for smaller systems, owing to their reduced size creating smaller recombination cross sections with re-colliding electron wave packets (with C$_{60}$ posing an anomalous exception to this rule). Our predictions pin-point fullerenes as an ideal system for exploring multi-electron interactions in HHG and strong-field physics, and should motivate experiments. 
\end{abstract}

\maketitle


\section{\label{sec:level1}Introduction}
High harmonic generation (HHG) is an ultrafast quantum process in which molecules are irradiated by an intense laser pulse and emit coherent radiation in multiples of the fundamental frequency of the driving laser \cite{Ferenc1}. This creates a broadband coherent emission that may reach up to KeV range and is useful for imaging and attosecond pulse sources \cite{KeV2,KeV3Gao}, as well as for HHG-based ultrafast spectroscopy \cite{Vozzi4_OrbTom,Farell5_2011,Borrego6_2022,Kneller7_2025,Marangos8_2016,Haessler9_2011,OferPRX10_2019,Gaarde11_2022,Ofer12_PRR2020,Ofer13_sciadv.2023,Worner14_PhysRevX.2018,Ofer15_PRL.2019}. HHG in gas-phase atoms and molecules is well understood by a mechanistic semi-classical trajectory picture for the electron dynamics \cite{Corkum18_PRL.1993,Lewenstein19_PRA1994}. In this framework, the electron is (i) tunnel-ionized by interaction with the laser, (ii) accelerated semi-classically in the continuum by the laser, and (iii) recombines with its parent molecule to emit an HHG photon. In certain molecular systems and laser-matter regimes, one can find conditions where multiple-orbitals contribute to this mechanism, which has been employed for ultrafast spectroscopies (e.g. see Refs. \cite{Le.multiorb_2009,lin.multiorb_exp2012,ZHAI_multiorb_ellip2025,Subhadeep2025SciAdv}). Typically though, no more than a handful of orbitals are involved due to exponential suppression of tunnel ionization from deeper states \cite{Le.multiorb_2009,lin.multiorb_exp2012,ZHAI_multiorb_ellip2025,Subhadeep2025SciAdv}. Moreover, interactions between electrons are almost always negligible, except for unique cases in strongly correlated systems \cite{Gilary2006,DelasHeras2020,Kraus2018, Smirnova2009Nature,Patchkovskii2006PRL, Vozzi4_OrbTom}, near resonances \cite{Wahyutama2019,ishikawa_elec.corre.hhg2017,PhysRevA.103.043114,ofer2026_resonant.hhg,reso.hhg_lein2011,Facciala2016,Fareed2018, Ofer2024_jpcl}, or in specially prepared conditions \cite{madsen.elec.corr-driven2024,ishikawa_elec.corre.hhg2017,Ofer12_PRR2020}. Note that this greatly differs from HHG in solid-state systems \cite{angel_ellip_solids.2017,sato_2020.prr,Sato2025_rev.npj} where many electrons interact with the laser and correlations are more widespread \cite{e-e.solids_2022_alam,Uchida2022,PhysRevLett.121.097402,Freudenstein2022,ChangLee2024,Jensen2024a}.

Fullerenes are carbon-allotrope nanomaterials that have drawn extensive interest recently owing to their unique structural and electronic properties \cite{fei2007_e-ecorre_c20,Samo_h2onc60_2017,becker_c60_ioni_prl2006,c60-xray.diff_sciadv.2025,Subhadeep2025SciAdv}. They are prospective candidates in a wide variety of novel fundamental and technological applications, e.g. in plasmonics where they exhibit sharp plasmon resonances that arise from multi-electron interactions and interferences \cite{Ju1993,Matsko2021,Subhadeep2025SciAdv,becker_c60_ioni_prl2006,c60-xray.diff_sciadv.2025,fullerene.photocurrent_acs.jpcc.9b01439,fullerene_qubit_2021_npj}. They can also be used as tailor-made sensor and electron acceptors due to the structural cage and its stability \cite{c60_acceptors_charge_transfer_2018,c60.sensors_2025}. Earlier works in HHG have focused primarily on C\textsubscript{60}, for instance showing that emission originates from electronic density oscillations on its surface contributed by $\pi$ electrons. HHG from C\textsubscript{60} was also experimentally measured already in 2009 with an enhanced harmonic yield compared to that of bulk carbon \cite{c60_hhg_exp_prl2009}. However, no work to date explored angular and ellipticity dependence of HHG in C\textsubscript{60}, and certainly not compared structural trends across the fullerene family. The role of correlations and multi-orbital features under intense driving is also not fully understood. 

Here, we perform a vast \textit{ab-initio} study of HHG from fullerenes considering various laser parameters and molecular sizes and polymorphs. This allows us to track down trends across species. We show that HHG cutoffs have qualitatively close ranges for different system sizes, owing to similar ionization energy bandwidths. In all examined cases we find very strong multi-orbital effects with HHG being contributed from many active states (reflected by a failure of the single-active electron approximation, large orbital ionization rates from multiple states, and clear interference effects in time-resolved Gabor spectrograms). We further show that dynamical electron-electron interactions are crucial for obtaining quantitative HHG yields in both plateau and perturbative harmonics, meaning correlations strongly alter fullerene HHG. By studying ellipticity and orientation dependence across the family in 3D we find that larger fullerenes generally exhibit a wider ellipticity dependence (following larger recombination cross sections) and a weaker orientation dependence (which is a result of their enhanced isotropicity). Thus, molecular size and laser polarization can be used as control knobs to tune HHG emission. Our results suggest fullerenes as optimal molecules for testing multi-electron interactions and interferences in gas-phase HHG, which should motive experimental studies. 

\section{\label{sec:theory}Methodology}


Let us first describe our methodological approach. Nonlinear light-induced electron dynamics in fullerenes is studied by investigating time-evolution of the system under intense laser driving \textit{ab-initio} with time-dependent density functional theory (TDDFT) \cite{Ullrich2025}. We employ the local density approximation (LDA) for the exchange-correlation (XC) functional within the adiabatic approximation and further add a suitable self-interaction correction (SIC) term to the XC functional \cite{SIC_2002}. The resulting electron dynamics follow the time-dependent Kohn-Sham (KS) equations, given as (in atomic units and length gauge)-
\begin{equation}
\begin{aligned}
i \partial_t \ket{\psi^{KS}_i(\mathbf{r},t)} &=
- \frac{1}{2} \nabla^2 \ket{\psi^{KS}_i(\mathbf{r},t)} \\
&+ \left[
V_{KS}(\mathbf{r},t)
- \mathbf{E}(t)\cdot\mathbf{r}
\right]\ket{\psi^{KS}_i(\mathbf{r},t)}.
\end{aligned}
\label{1}
\end{equation}
The KS potential $V_{KS}(\mathbf{r},t)$ can be decomposed as: 
\begin{equation}
V_{KS}(\mathbf{r},t)=V_{ion}+\int d^3r' \frac{\rho(\mathbf{r'},t)}{|\mathbf{r}-\mathbf{r'}|}+V_{XC}[\rho(\mathbf{r},t)]. \label{2}
\end{equation}
Here, $\ket{\psi^{KS}_i(\mathbf{r},t)}$ is the $i^{th}$ time-dependent KS orbital. Further, $V_{ion}$ accounts for the electron interaction with the nucleus and core electrons, the second term accounts for the Hartree potential where $\rho(\mathbf{r},t)$ is the time-dependent electron density, and $V_{XC}[\rho(\mathbf{r},t)]$ denotes the XC potential.
$\mathbf{E}(t)$ in equation \eqref{1} is the time-dependent electric field of the driving laser pulse. It reads within the dipole approximation as:  $\mathbf{E}(t) = E_{0} f(t) cos(\omega t) \hat{e}$. Here $E_{0}$ is the field amplitude, $f(t)$ is the envelope function which has a $sin^2$ form, $\omega$ is the carrier frequency, and $\hat{e}$ is a normalized unit vector that represents the laser polarization (either linear in different directions in space, or elliptical with varying ellipticity within different planes). Unless stated otherwise, the driving laser field was chosen as an 8-cycle long pulse with a peak intensity of 10\textsuperscript{14} W/cm\textsuperscript{2}.

Ion motion is neglected within the fixed-nuclei approximation. In order to avoid unphysical reflections of electron from the simulation box boundaries, we implement a complex absorbing potential (CAP) during the time-evolution. The induced non-linear polarization is evaluated from the time-dependent electron density as $\mathbf{P}(t) = \int d^3r\mathbf{r}\rho(\mathbf{r},t)$, where the density $\rho(\textbf{r},t)$ is obtained from the time-dependent KS orbitals directly. The HHG spectrum is obtained from $\mathbf{P}(t)$ by taking Fourier transform of the dipole acceleration and filtration by the pulse envelope, and unless stated otherwise is taken as a sum of all emission polarizations.  The geometrical centers of the molecules are fixed at the 
origin of the simulation box (sphere shaped) and laser polarizations are considered with respect to this simulation box-fixed reference frame.

All DFT and TDDFT calculations were performed using the Octopus code \cite{octopus2020}. The optimized geometries of fullerenes were obtained from \textit{ab-initio} quantum chemical calculations using Gaussian program \cite{g16_ref}, where we employed default optimization criteria by the program within the DFT method. The optimized structures were tested by subsequently performing vibrational frequency analysis where no imaginary frequencies were found. 
The KS equations were discretized on a Cartesian grid with a spherical simulation box. The radius of the box converged at 50 a.u., which facilitates appropriate spatial dimensions for the laser-driven electronic motion and its subsequent absorption. The selected grid spacing is 0.3 Bohr for all molecules. The ground-state KS equations are solved self-consistently for each fullerene with a self-consistent field (SCF) threshold of 10$^{-8}$ Hartree. We employed the average-density SIC to obtain correct long-range asymptotic behavior of the XC potential \cite{SIC_2002} and consequently improved ionization energies. For the core states, we employ frozen core approximation using norm-conserving pseudopotentials \cite{norm-cons_pseudo_1998}. The calculated ionization energies (in terms of ground state KS eigenvalues) for HOMO and innermost orbitals for the fullerene family are provided in Table \ref{tab:Ip_energies}.

\begin{table}[h!]
\caption{Ionization energies for the HOMO and innermost valence orbitals (in a.u.) for various fullerenes.}
\label{tab:Ip_energies}
\centering
\begin{tabular}{lcc}
\hline
System & HOMO  & Innermost  \\
\hline
C$_{20}$ cage & 0.268971 & 1.040804 \\
C$_{20}$ bowl & 0.318820 & 0.957253 \\
C$_{20}$ ring & 0.250912 & 0.819954 \\
C$_{24}$      & 0.292205 & 1.003581 \\
C$_{30}$      & 0.261558 & 1.006326 \\
C$_{36}$      & 0.270880 & 0.985806 \\
C$_{40}$      & 0.278019 & 0.981173 \\
C$_{60}$      & 0.273401 & 0.951162 \\
\hline
\end{tabular}
\end{table}

After GS diagonalization, we carried out TDDFT simulations by propagating all occupied KS orbitals with a Lanczos propagator and a converged time step of 0.2 a.u. A CAP width is converged at 20 a.u. In all simulations the initial state prior to laser excitation was chosen as the ground state with all occupied orbitals up to the HOMO.

\section{\label{sec:results}Results and Discussions}
 
\subsection{\label{sec:theory_gs1}HHG structure and multi-electron effects}

\begin{figure}
\centering
\includegraphics[width=1.0\linewidth]{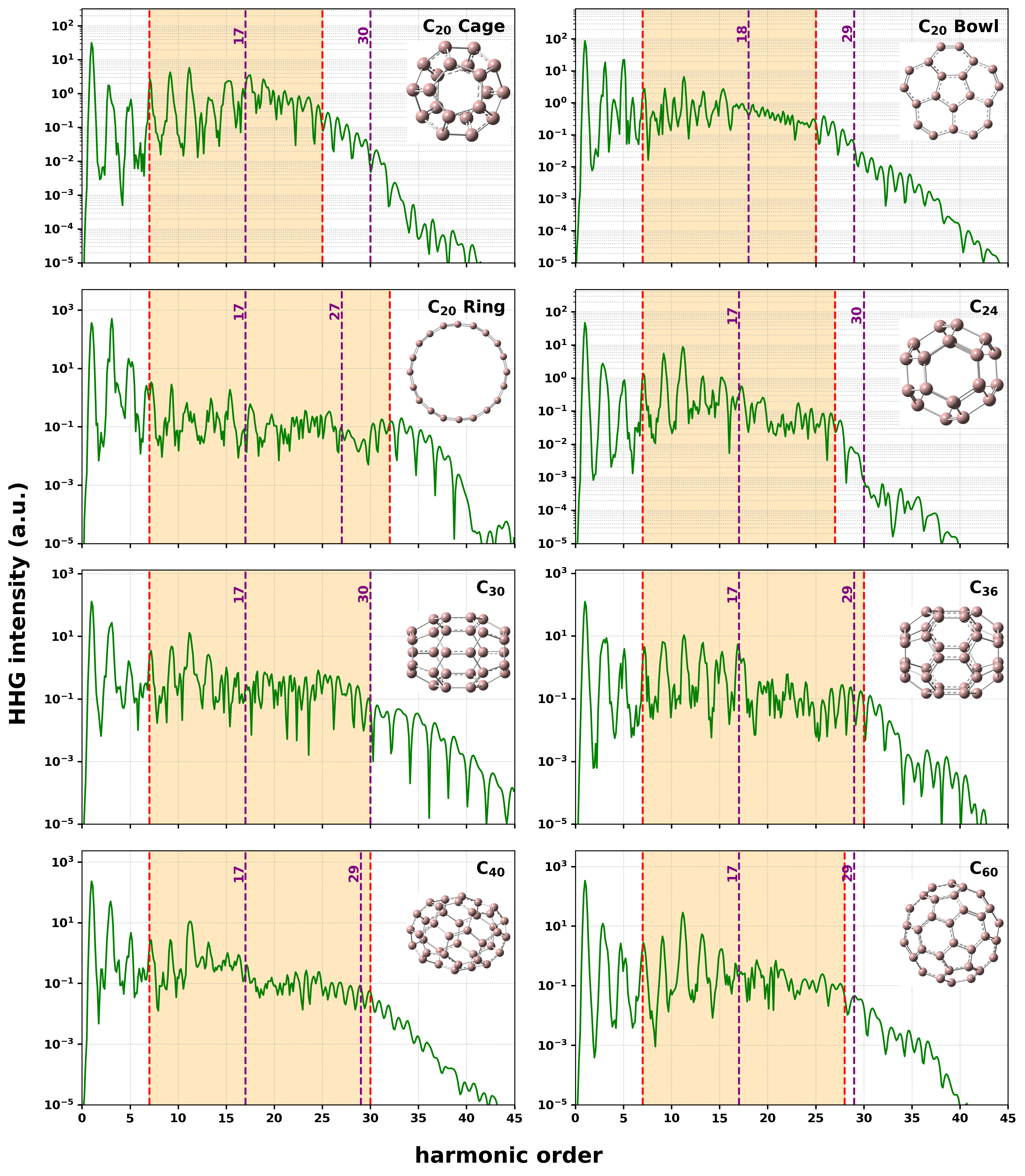}%
\caption{Harmonic generation from fullerenes family (as they progressively increase in size) for the driving laser linearly polarized along x-axis. The geometry of each fullerene is shown in inset. The plateau is marked by a background orange coloring, with dashed purple lines indicating typical semi-classical harmonic cutoffs from the HOMO and deepest valence orbital in the system (with HO indicated numerically).}
\label{fig:1}
\end{figure}

We begin by numerically exploring the typical HHG emission from different size and polymorphs of fullerenes. Figure \ref{fig:1} shows HHG spectra from fullerenes increasing in size. We notice a universal HHG emission pattern: a similar perturbative region, followed by a wide spanning plateau region across $\sim$21 harmonic orders (HOs) with typical range from HO 7 to 30 (with slight variations). Spectra show only modest modifications in the HHG yield between species. From a semi-classical trajectory perspective \cite{Lewenstein19_PRA1994,Corkum18_PRL.1993}, the HHG cut-off corresponding to HOMO orbital (first vertical line in Fig. \ref{fig:1} denotes the ionization potential, and first purple dashed line denotes the cutoff from the HOMO) is very close for all systems due to similar ionization potentials. Similarly, the harmonic cutoff corresponding to the deepest valence orbital is also close across species (denoted in Fig. \ref{fig:1} by the second purple dashed line). Indeed, Table 1 \ref{tab:Ip_energies} enlists HOMO and the deepest valence state ionization energies of fullerenes, showing that orbital energies have similar ranges across systems that appears to form an energy-band like electronic structure and similar bond hybridization and chemical nature. This structure in turn is behind the universal HHG emission ranges in Fig. \ref{fig:1}. Interestingly, as the system size increases we notice that the plateau extends more and more beyond the cutoff associated with the HOMO, and closer to the cutoff expected by the deepest valence state, indicating successively increased orbital contributions from deeper states.   

\begin{figure}
\centering
\includegraphics[width=1.0\linewidth]{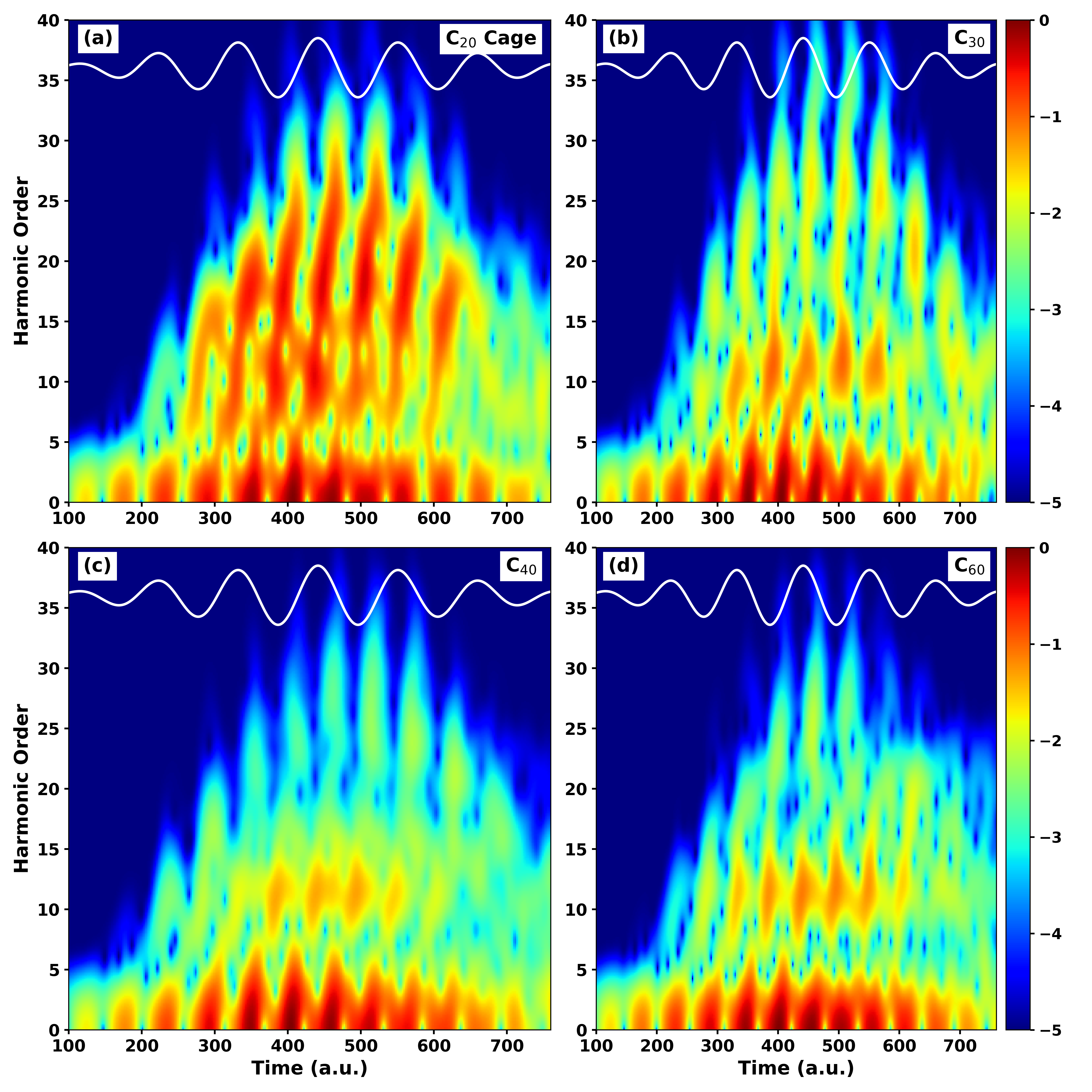}%
\caption{Gabor time-frequency analysis for selected fullerenes. The driving laser pulse is overlaid along each panel to elucidate underlying electron dynamics of harmonic emission for each fullerene, which also showcases their relative sub-cycle dependence, harmonic intensity and emission windows. The laser driving field is taken similar to Figure \ref{fig:1}. The cage fullerene size is indicated in the top-right of each panel.}
\label{fig:2}
\end{figure}

We next explore the time-resolved dynamics of the HHG process through a Gabor transform (a time-resolved Fourier analysis). Figure \ref{fig:2} shows Gabor spectrograms of C\textsubscript{20}, C\textsubscript{30}, C\textsubscript{40}, and C\textsubscript{60}. Dominantly short trajectories can be observed with a clear attochirp, as expected in 3D simulations. From Fig. \ref{fig:1} we observe interference patterns beyond HO 17, i.e. after the HHG cutoff associated with the HOMO, which continues up to the semi-classical cutoff corresponding to the deepest valence orbital (second vertical purple lines in Figure \ref{fig:1}). These interferences and cutoff ranges are clear indication of multi-orbital contributions to HHG emission from fullerenes. 

Turning to the ground state calculations, it is evident that fullerenes have closely-spaced energy level distributions that form an equivalent of energy bands (similar to a solid-state system with periodic boundaries along a sphere). Moreover, due to the high symmetry and non-abelian point group of the species, multiple energy-degenerate orbitals exist across systems. Such multiple contributions in a tight energy window should lead to strong interference effects. Notably, HHG interferences indeed become more severe for larger fullerene sizes (see Fig. \ref{fig:1} and \ref{fig:2} around the plateau region). This reflects the higher orbital densities within the same energy window. That is, since the innermost and outermost orbitals have roughly similar ionization energies across systems, then the fact larger molecules have many more states means a much larger density of states per unit energy. This in turn translates to more dense bands and noticeable multi-electron interferences in HHG and time-resolved spectrograms.


\begin{figure}
\centering
\includegraphics[width=1.0\linewidth]{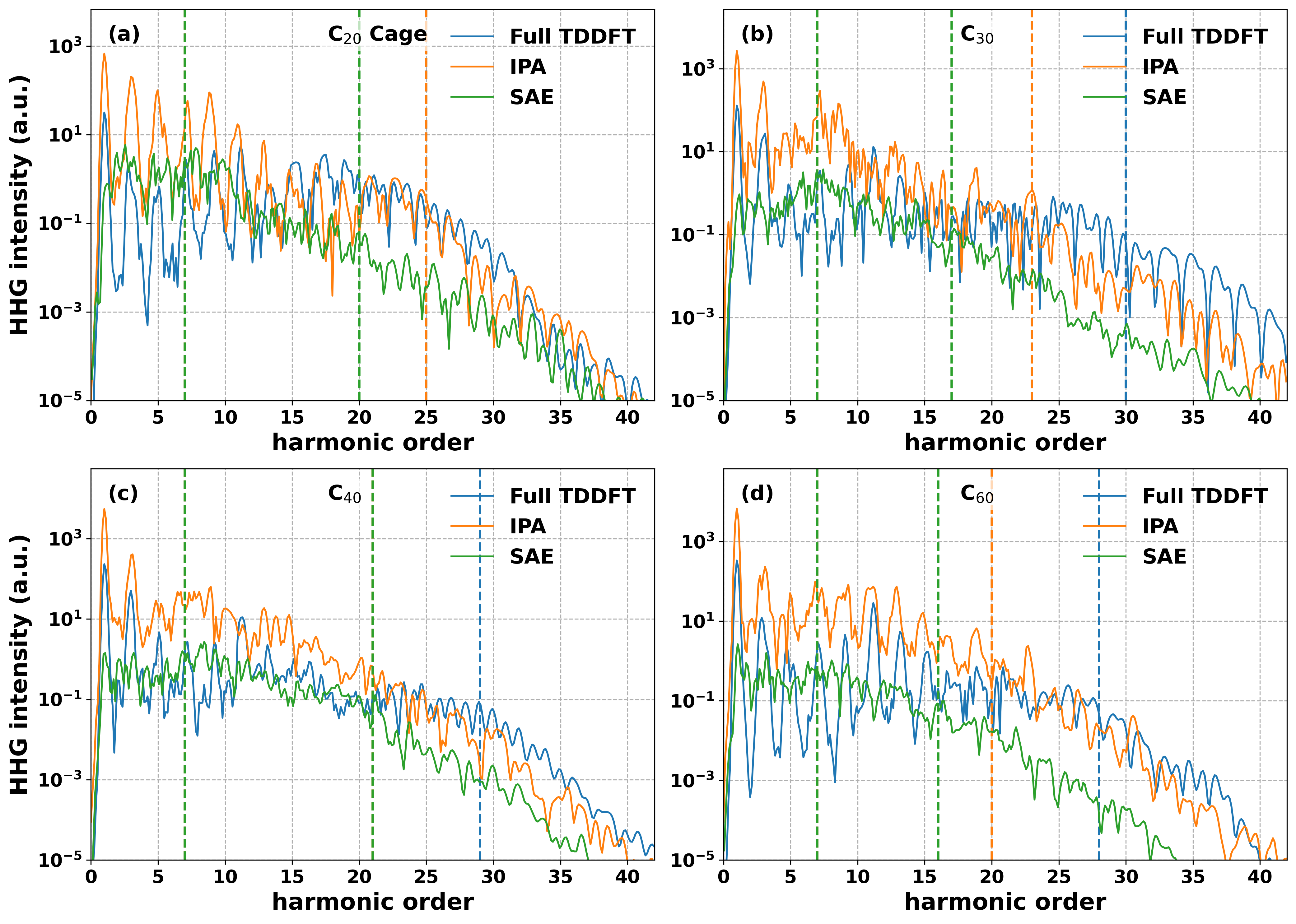}%
\caption{Illustration of roles of electronic correlation and multi-orbitals contribution to the HHG emission process for the selected fullerenes. The correlated electron dynamics is studied employing full TDDFT simulations (blue), whereas the correlated dynamics is kept frozen during time-evolution of the system IPA (orange). Within the SAE, only the HOMO KS orbital is allowed to time-evolve (green). The vertical lines in each panel present plateau regions for each respective case.}
\label{fig:3}
\end{figure}

To further explore this effect, we compare for exemplary C\textsubscript{20} system the HHG emission calculated within the single active electron (SAE) approximation (Fig. \ref{fig:3}(a)). From Fig. \ref{fig:3}(a), we clearly see that the SAE fails terribly. Besides incorrectly describing the cutoff energy, the HHG yields themselves are also poorly described compared to the full TDDFT simulation. Figure \ref{fig:3} show similar features in other fullerene sizes as well, indicating a general trend. We further calculated the KS orbital populations during laser driving, showing that indeed multiple orbitals substantially contribute to ionized states (see SI). This result undeniably supports the evidence above and validates involvement of multiple orbitals in the ultrafast dynamics and HHG. 

Lastly for this section, we study the role of electronic correlation in the ultrafast electron dynamics. We consider two distinct levels of theory: (i) Full TDDFT where electrons dynamically interact via electron density changes that modulate interactions over time via the Hartree and XC terms. (ii) Restricting the KS potential to be temporally frozen to its initial form, leading to the independent particle approximation (IPA). In the IPA, electrons interact with the mean-field potential of the ground state system, but correlations do not dynamically evolve in the sense that the dynamics can be captured by the static KS potential acting as a non-interacting model. 

Typically, the IPA is an extremely good approximation for HHG, which is indicative of the weakly-correlated dynamics induced by the laser and the fact that the laser-matter interaction term dominates the dynamics. In fullerenes, however, we expect potentially stronger features of interactions as recently measured in photoemission \cite{Subhadeep2025SciAdv}, which is a result of strong multi-electron plasmonic resonances. Figure \ref{fig:3} shows comparison across different cage systems between TDDFT and IPA, indicating the the IPA generally fails. Indeed, the IPA correctly predicts HHG cutoff energies, but fails to predict the harmonic structure in both the plateau and in the perturbative region of low-order harmonics. In particular, we find that the IPA overestimates HHG yields for lower HOs and often leads to noisier spectra. This effect is likely a result of the involvement of multiple orbitals in the emission process as discussed above, which re-normalizes the KS potential and Hamiltonian as the orbitals ionize on femtosecond timescales (an effect that is neglected in the IPA). In that respect, our simulations pin-point fullerenes under intense laser fields as potentially ideal systems to explore dynamical many-body effects in gas phase molecules.

\subsection{\label{sec:theory_gs}HHG angular and ellipticity dependence}

\begin{figure}
\centering
\includegraphics[width=1.0\linewidth]{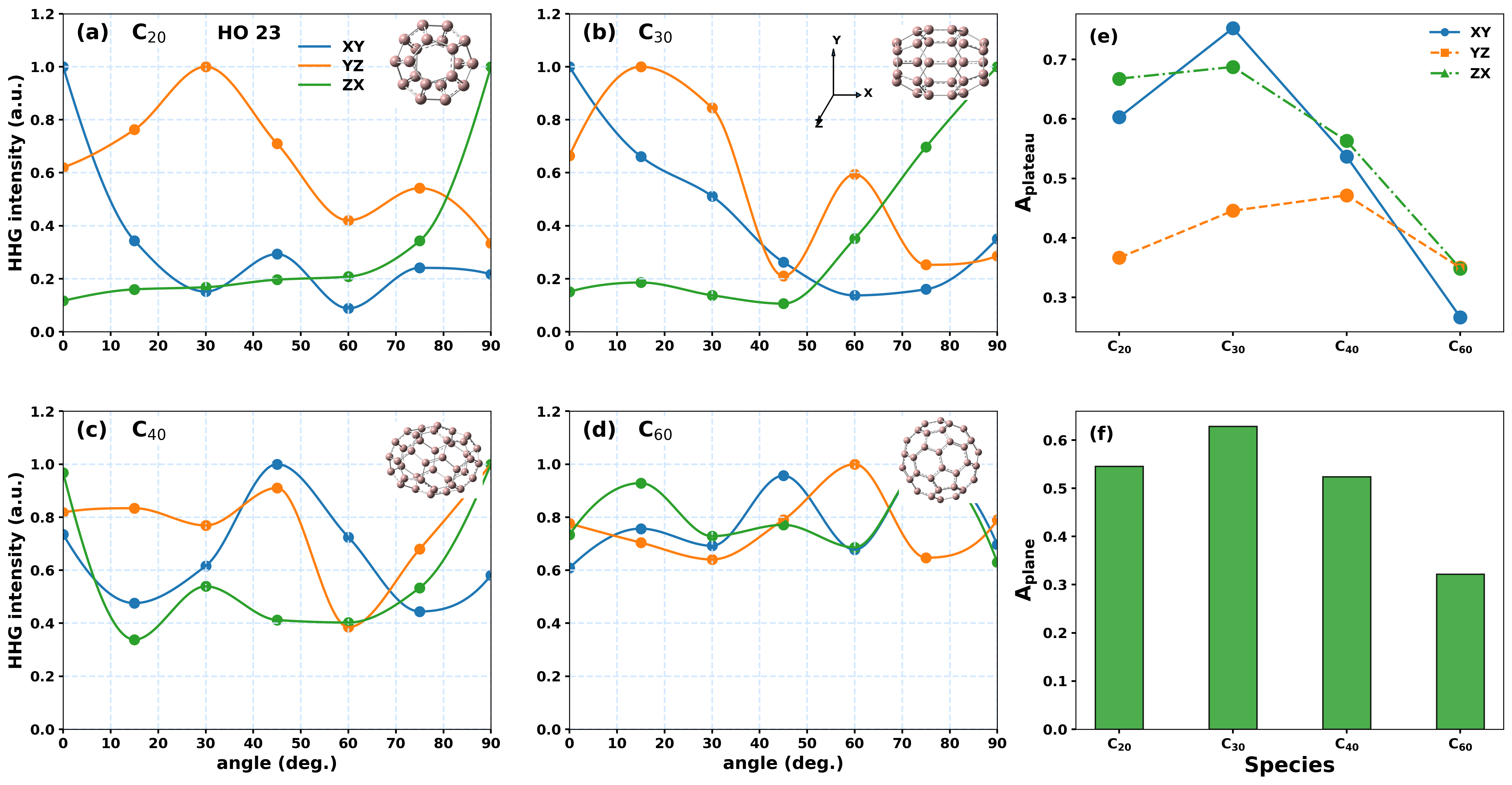}%
\caption{(a-d) Harmonic yield dependence on driving laser angular orientation for a select HO 23 (within the plateau) across three orthogonal main planes, and for different size fullerenes (shown in inset). Simulations performed in similar conditions to Fig. \ref{fig:1}. (e-f) Anisotropy parameter averaged over the plateau across different planes, (A\textsubscript{plateau}). (f) total anisotropy parameter averaged over plateau harmonics and three main planes, (A\textsubscript{plain}), quantifying total angular sensitivity for each fullerene.}
\label{fig:4}
\end{figure}

We next studied the angular dependence of HHG emission across the fullerene family. Note that throughout we consider an aligned molecule rather than a spatially-isotropic ensemble. Figure \ref{fig:4} presents exemplary angular dependence for a select harmonic in the plateau for fullerenes C\textsubscript{20}, C\textsubscript{30}, C\textsubscript{40} and C\textsubscript{60}, though we calculate such dependencies for all harmonics in the spectra (not presented). Generally, we noticed that the HHG yields vary differently with the laser driving angles in different fullerene planes, which can have stronger/weaker dependence, depending on conditions. In order to quantify these effects across molecular sizes, we define the HHG angular-dependence bandwidth. The bandwidth is given by the normalized difference between maximal and minimal HHG yields for a given harmonic order, and within a given molecular plane: XY, YZ, or ZX. This measure, describes an anisotropy parameter A\textsubscript{q} that quantifies the degree of angular dependence for a given order q, which would completely vanish in a fully isotropic system. Mathematically, it is given by: $A_q = \frac{I_q^{\max} - I_q^{\min}}{I_q^{\max} + I_q^{\min}}$; where $I_q$ shows intensity of the q\textsuperscript{th} harmonic in the plateau. To further get a quantitative measure we average this bandwidth across all plateau harmonics in each molecule (Fig. \ref{fig:4}(panel e)), yielding plateau-averaged but plane-resolved anisotropy parameter A\textsubscript{plateau}. Further averaging over three main planes (XY, YZ, and ZX) we obtain an anisotropy parameter A\textsubscript{plain}, providing a single scalar that quantify how strong a given species leads to angularly-dependent HHG (Fig. \ref{fig:4}(panel f)). Our results show that larger fullerenes exhibit a generally less sensitive angular dependence, which decreases linearly with size. 

This result is both intuitive, as well as counter-intuitive, in the sense that on the one hand large fullerenes are usually of a higher symmetry group. This naturally suggests that larger systems should be more isotropic and exhibit less angularly-sensitive responses. On the other hand, we recall that in practice larger systems exhibit a higher density of molecular bonds per unit angle, and even though the molecular point group increases in size with the molecule and contains more symmetry elements, the highest-order elements remain fixed and are C$_5$ rotations. Ultimately, the former wins over in the HHG angular sensitivity and indeed larger systems have a more isotropic uniform response.

\begin{figure}
\centering
\includegraphics[width=1.0\linewidth]{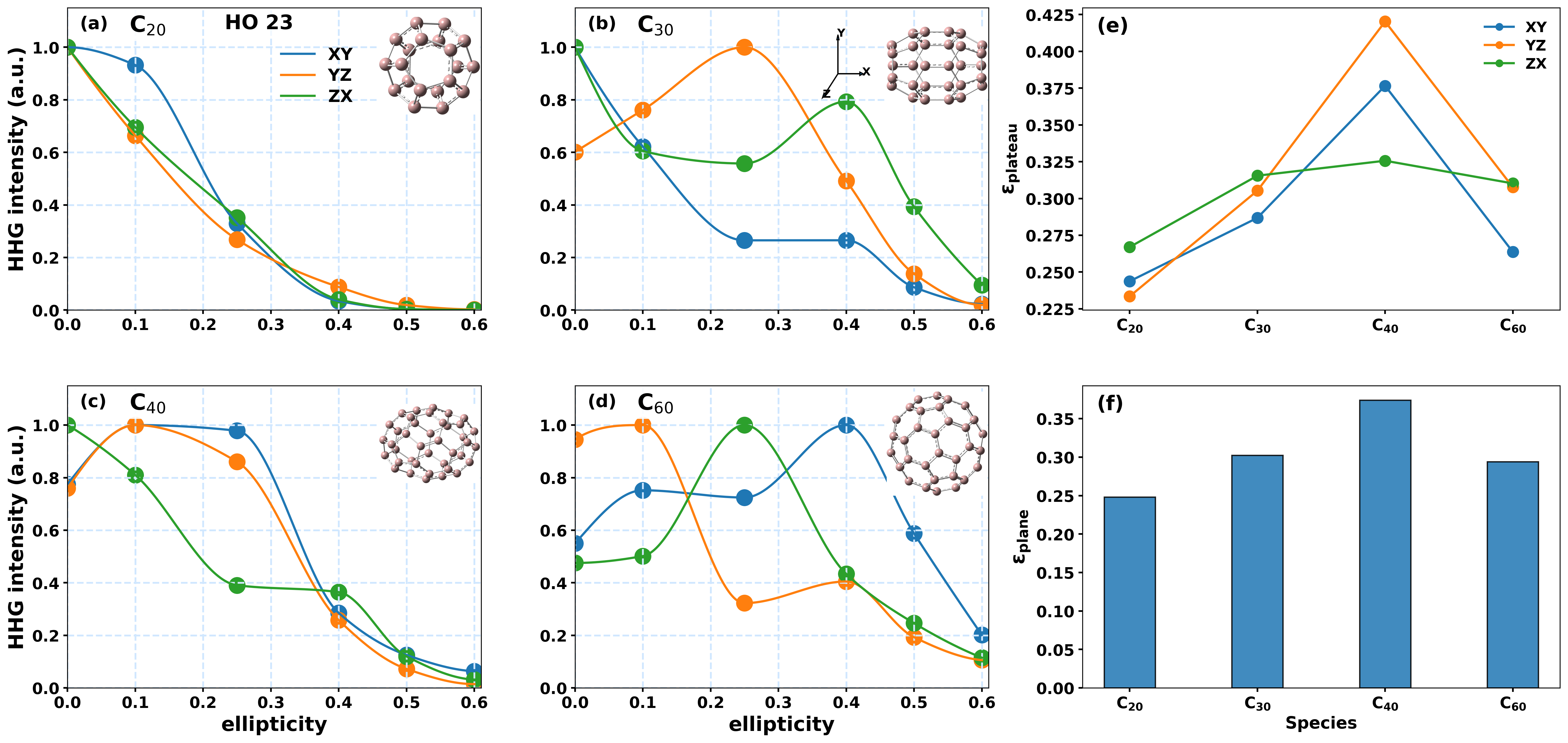}%
\caption{(a-d) Ellipticity dependence of a selected harmonic HO 23 (within the plateau) in different size fullerenes, calculated in similar conditions to Fig. \ref{fig:1} but with elliptical driving. (e) Plateau-averaged fitted width for the exponentially-decaying ellipticity sensitivity in each system and plane. (f) Plane and plateau averaged width of exponentially-decaying ellipticity sensitivity in each system.}
\label{fig:5}
\end{figure}

At the next stage, we repeat this rigorous analysis but for HHG ellipticity dependence. In atomic gases HHG is well established to decay exponentially with the driving laser ellipticity \cite{ellipMoller_2012,ellip_lin2021} (unlike in solids \cite{Ghimire2019} and liquids \cite{ofer2026_liq}). This is a result of the delocalization of the ionized electron, which correspondingly misses the parent atom and reduces recombination cross sections. In molecules some re-collision for small ellipticity in the driving laser is attainable, but typically there is still a strong decay of the HHG yields vs $\varepsilon$. For oriented molecules one can generally get non-trivial dependencies \cite{ellip_mole_2007} just like has been shown in condensed phases \cite{Yoshikawa2017,You2017,Herling2026,Heide2022a,Neufeld2023a}, which arises from multiple scattering and recombination centers. Figure \ref{fig:5} shows our main result for one select harmonic order (HO 23) in the plateau of each fullerene, which illustrates broad and complex ellipticity dependence (other orders are not shown but analyzed together below). This broad and highly oscillatory dependence reflects the complex nature of the system and the multiple potential ionization and recombination sites available for recombining electrons. Still, there is a strong decay with $\varepsilon$ such that in all cases the HHG yield vanishes above $\varepsilon\sim0.6$. We fit the normalized yield per harmonic, per molecular plane, to a decaying Gaussian, and obtain the relative width of the Gaussian, $\varepsilon_0$. By averaging this fitted quantity across all harmonic orders in the plateau of each species we obtain Fig. \ref{fig:5}(e). Further averaging over the major planes (XY, YZ, ZX), we obtain Fig. \ref{fig:5}(f), which shows the typical ellipticity emission width of each system. 

Two main key features arise. First, for larger fullerenes broader ellipticity dependence is obtained. This is an expected result considering the semi-classical dynamics of electron trajectories - larger systems pose larger targets with higher cross sections for re-collision. Second, we note that C\textsubscript{60} is an anomaly in this trend. We hypothesize that this arises from its complex plasmonic resonance multi-electron contributions to HHG (discussed above). Such interactions could lead to advanced interferences that alter the width of the ellipticity dependence, which the Gaussian fit anyways approximates very roughly.

\section{\label{sec:summary}Summary}
To summarize, we numerically explored HHG from fullerenes C\textsubscript{20} through C\textsubscript{60}, including isomers, with \textit{ab-initio} TDDFT simulations. Our predictions pin-point gas-phase fullerenes as optimal systems for exploring light-driven electronic correlations in strong-field physics. Correlations are seen to generally reduce HHG yields and cause emission of sharper and cleaner harmonics peaks. We further showed that across the fullerene family multi-orbital features naturally arise, while the single active electron picture is inapplicable. This leads to orbital interference effects in HHG plateaus and time-resolved spectrograms. The harmonic spectra was otherwise shown to have relatively similar plateau and cutoff features across molecular sizes, which is a result of the rather uniform orbital energy ranges in all systems. We further analyzed HHG ellipticity and angular dependence vs molecular size, and uncovered that larger molecules have a reduced sensitivity to the driving laser ellipticity, as well as a reduced sensitivity to the driving laser angle. Both effects correspond to expected semi-classical physics \cite{ellip_mole_2007,ellipMoller_2012,ellip_lin2021}, apart from an anomaly for C$_{60}$ that has a strong than expected decay rate with the driving laser ellipticity (which might arise from stronger plasmonic resonances \cite{Subhadeep2025SciAdv}).

Looking ahead, our results should motivate further exploration in both experiment and theory of strong-field driven phenomena from fullerenes. We expect that phenomena related to HHG such as charge migration \cite{CM2025_chemSci,pnas2025_nikolay,Cederbaum1999,Mansson2021,Worner2017,Calegari2014Science} and photoemission \cite{PIcarbon_sciadv.2023,Subhadeep2025SciAdv,vitali_PI2023,Wanie2024,Neufeld2025} will similarly be sensitive to many-body features. Our work also motivates searching for connections between fullerenes and solid-state carbon systems such as graphene and diamond \cite{Yoshikawa2017,Chen2025a,graphenehhg_Cox2017,diamondhhg_technion2026}, which have recently been explored in HHG and share similar chemical attributes.

\section{\label{sec:acknowledgments}Acknowledgments}
O.N. gratefully acknowledges the scientific support of Prof. Dr. Angel Rubio and the Young Faculty Award from the National Quantum Science and Technology program of Israel’s Council of Higher Education Planning and Budgeting Committee. O.N. and K.A.D. gratefully thank the Technion Helen Diller Quantum Center for financial support.


\bibliography{manus}

\clearpage
\onecolumngrid

\appendix
\renewcommand{\thefigure}{S\arabic{figure}}
\setcounter{figure}{0}
\title{Supplementary Information (SI)}
This supplementary information (SI) includes additional technical details showcasing convergence of our numerical simulations carried out in the main text and some additional results, as well as complementary results.

\section{S1: Numerical Convergence}
Here we present convergence tests for the selected time step, complex absorbing potential (CAP) width; denoted as CAP width and simulation box size. We select C$_{20}$ as the representative fullerene molecule to showcase these convergences. 
\subsection{Time step convergence}
\begin{figure}
    \centering
    \includegraphics[width=1.0\linewidth]{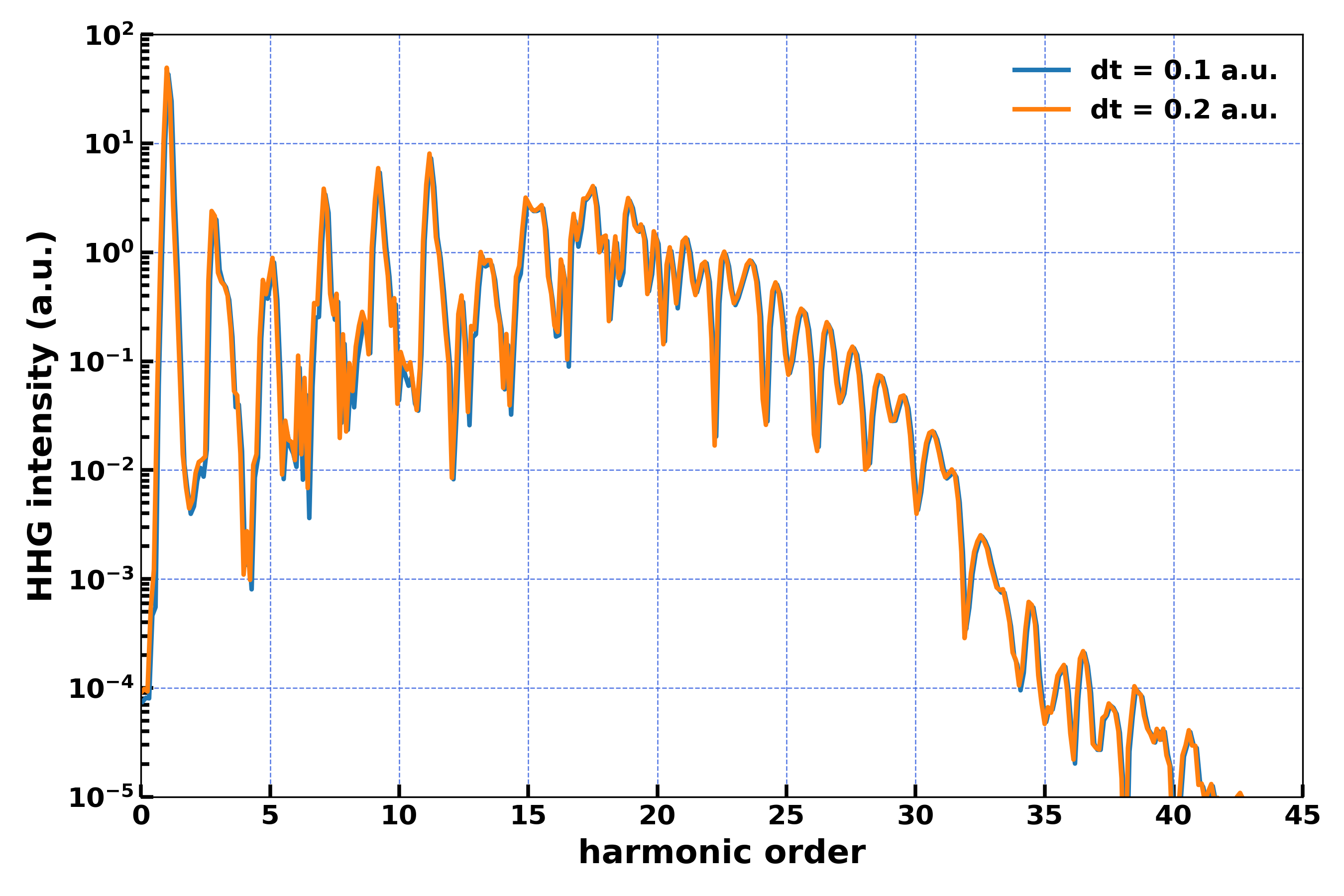}
    \caption{HHG dependence on the selected time step in the present \textit{ab-initio} simulations. The convergence is tested against time steps of 0.1 and 0.2 a.u.}
    \label{fig:S1}
\end{figure}
A time step of 0.2 a.u. was chosen for time-dependent simulations in our work. We tested this time-step against a smaller step size of 0.1 a.u. and found that the simulations are overlapped all over the harmonic range for both the cases. We illustrate this in the Fig. \ref{fig:S1}. 
\subsection{CAP width convergence}
\begin{figure}
    \centering
    \includegraphics[width=1.0\linewidth]{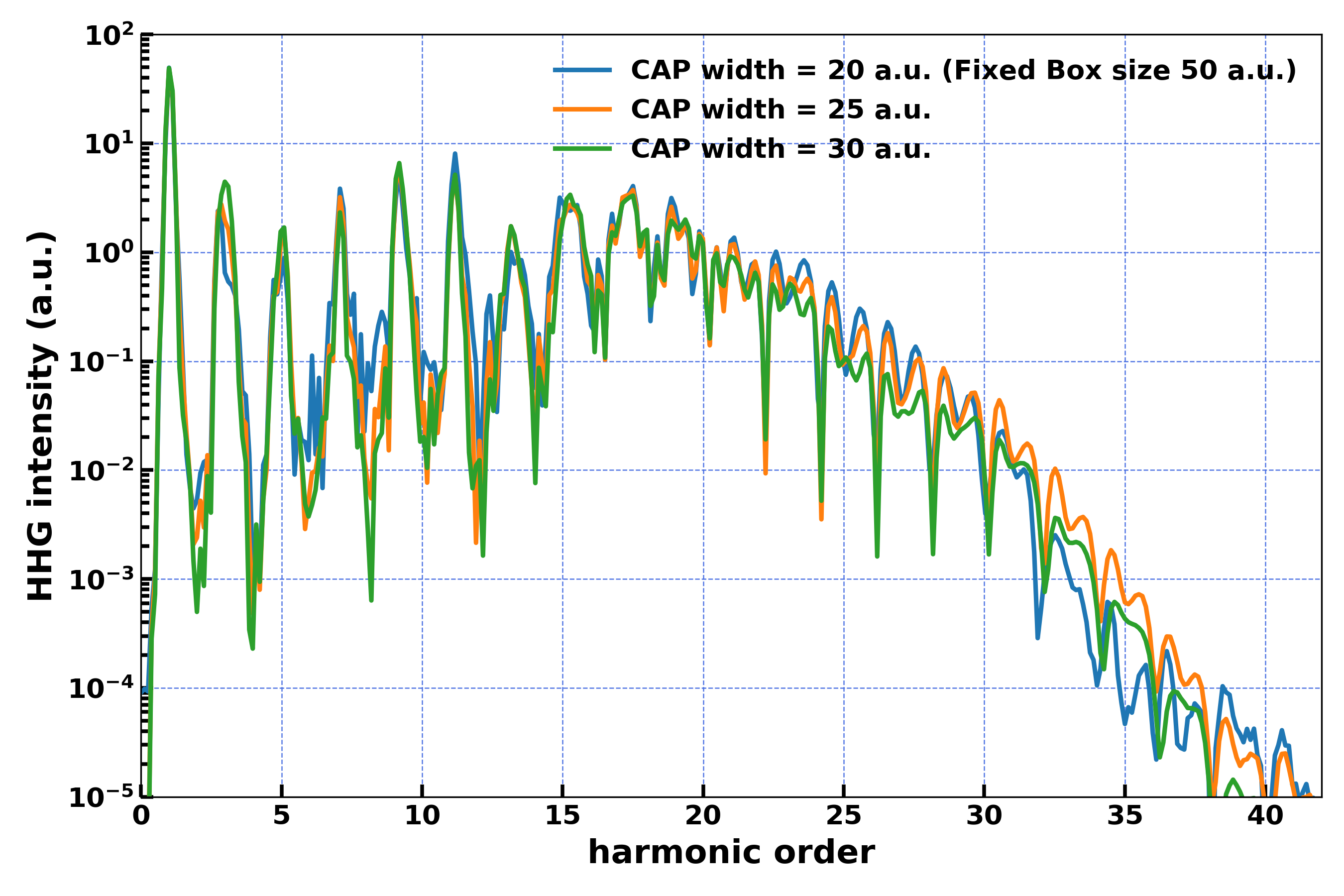}
    \caption{HHG dependence on the selected CAP width in the present simulations. As we successively increase the CAP Width keeping simulation box radius fixed, harmonic signals do not change.}
    \label{fig:S2}
\end{figure}
Fixing the simulation box radius, we increase the CAP width starting from 20 a.u. and successively increase it to illustrate that there is no loss/accumulation of harmonic signal due to inappropriate width of absorbing potential during the time-dependent propagation of the electron wavepacket. The convergence is testified via Fig. \ref{fig:S2}.
\subsection{Simulation box size convergence}
\begin{figure}
    \centering
    \includegraphics[width=1.0\linewidth]{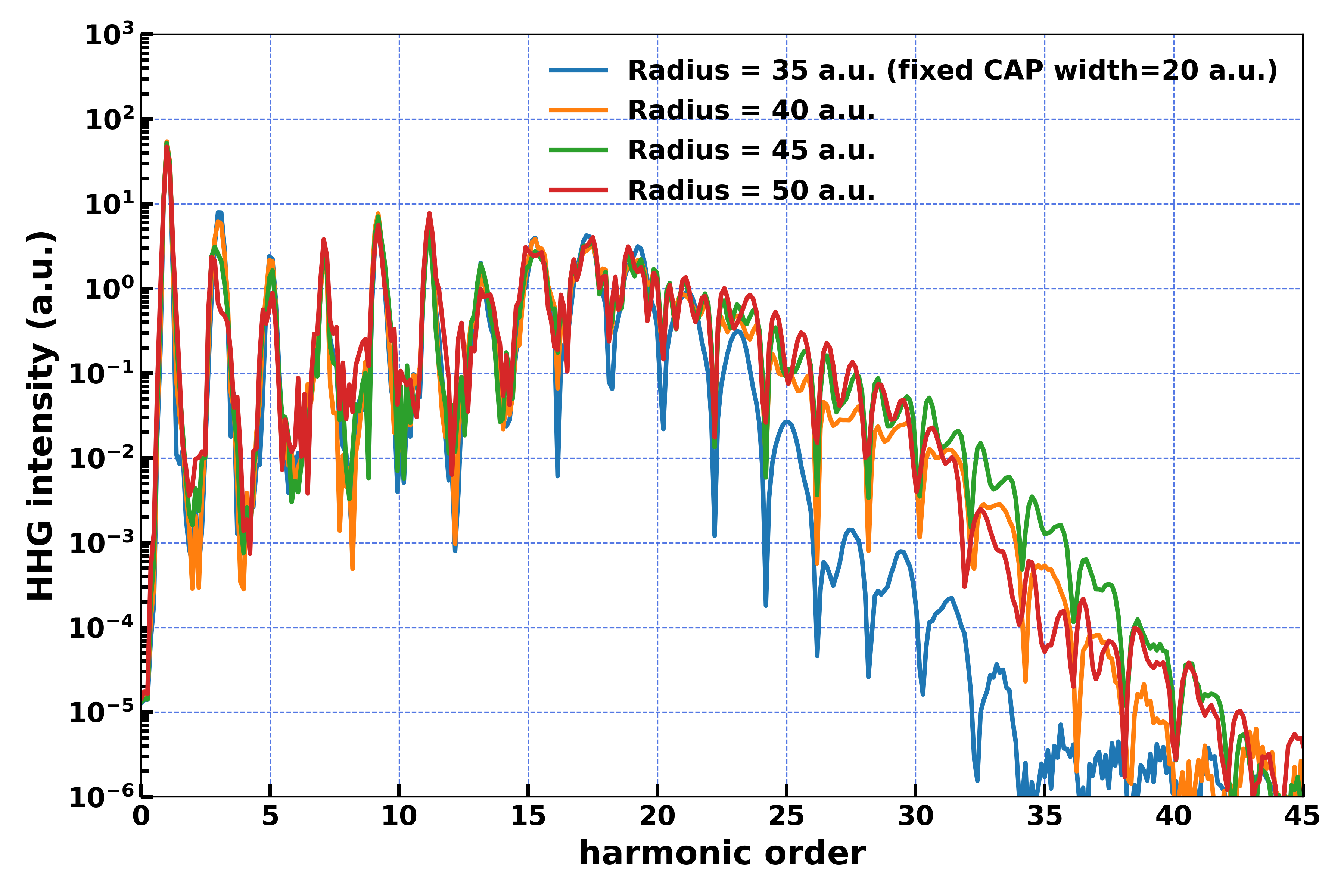}
    \caption{HHG dependence on the selected simulation sphere radius, while fixing the width of the CAP. As we successively increase the radius the HHG spectra converges.}
    \label{fig:S3}
\end{figure}

Next, we test the convergence of the selected simulation box size as shown in Fig. \ref{fig:S3}. We do so by keeping CAP width fixed at its converged value of 20 a.u. and increasing the box size from 35 a.u. until 50 a.u. It is evident that the simulation box choice of 50 a.u. is a converged size. Simulations in main text employed the strictest radius of 50 Bohr to ensure stability.
\subsection{Combined convergence of simulation box size and CAP Width}
\begin{figure}
    \centering
    \includegraphics[width=1.0\linewidth]{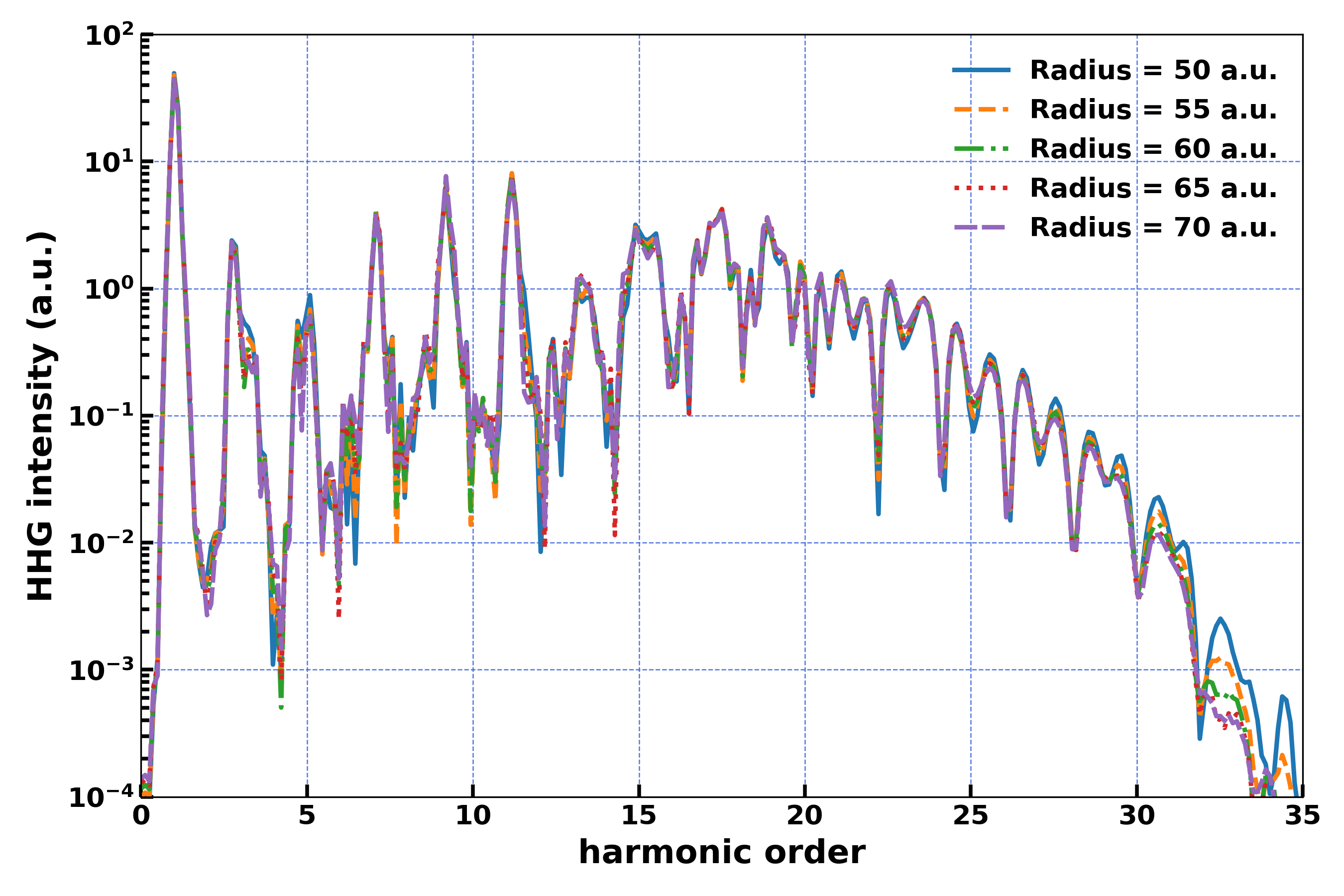}
    \caption{HHG dependence on the selected CAP width while simultaneously increasing the simulation box radius in order to make sure the CAP width is converged (see description in text).}
    \label{fig:S4}
\end{figure}
Next we simultaneously converge with respect to both the simulation box size and CAP width in the following manner: we increase the simulation sphere radius successively from 50 a.u. to 70 a.u., meanwhile enlarging the CAP width in the similar amount as the increment in box radius beginning from a CAP width of 20 a.u. at box radius 50 a.u. (which fixes the dynamical region where the electronic wave function can live but successively increases the CAP region) Fig. \ref{fig:S4} confirms that the selected box radius 50 a.u. and the corresponding CAP width of 20 a.u. is sufficient for accurately simulating the HHG dynamics of fullerenes with otherwise minor changes in the plateau region.

\section{Additional Results}

\begin{figure}
    \centering
    \includegraphics[width=1.0\linewidth]{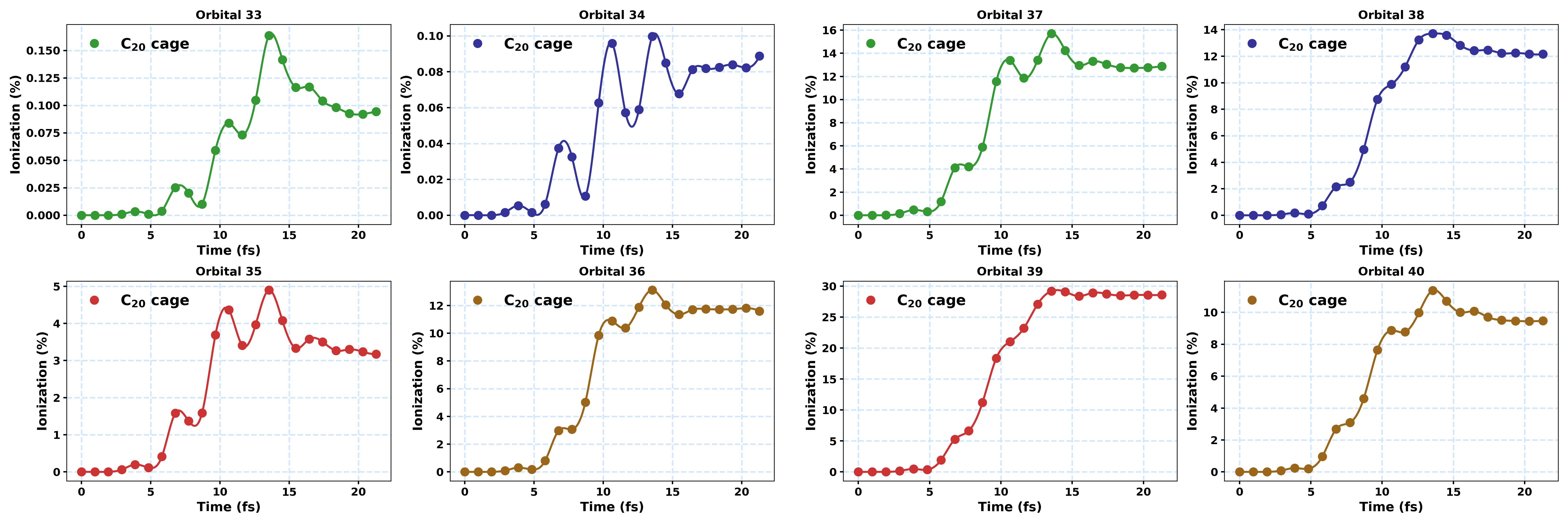}
    \caption{KS orbital ionization dynamics in C$_{20}$ cage in similar conditions to those studied in the main text. The results clearly indicate multiple orbitals ionized during the dynamics, with substantial contributions still arising down to HOMO-7 and more (not shown). Here the 40\textsuperscript{th} orbital denotes the HOMO.}
    \label{fig:S5}
\end{figure}

We further analyze the time-dependent occupations of individual KS states during laser driven dynamics in Fig. \ref{fig:S5}. Occupations are calculated by directly projecting the time-dependent KS orbitals onto those of the ground state at $t=0$. This complements the analysis in the main text discussing multi-orbital contributions. Fig. \ref{fig:S5} shows that indeed multiple orbitals ionize during interactions with the laser (in C$_{20}$ cage as an exemplary case), which supports the main text conclusions of multi-orbital effects also dominating the HHG response. 

\end{document}